# Quantum advantage in a molecular spintronic engine that harvests thermal fluctuation energy

B. Chowrira[1,2*], L. Kandpal[1*], M. Lamblin[1], F. Ngassam[1], C.-A. Kouakou[1], T. Zafar[1], D. Mertz[1], B. Vileno[5], C. Kieber[1], G. Versini[1], B. Gobaut[1], L. Joly[1], T. Ferté[1], E. Monteblanco[3], A. Bahouka[4], R. Bernard[1], S. Mohapatra[1], H. Prima Garcia[6], S. Elidrissi[6], M. Gavara[6], E. Sternitzky[1], V. Da Costa[1], M. Hehn[3], F. Montaigne[3], F. Choueikani[2], P. Ohresser[2], D. Lacour[3], W. Weber[1], S. Boukari[1], M. Alouani[1], M. Bowen[1@]

[1] *Institut de Physique et Chimie des Matériaux de Strasbourg, UMR 7504 CNRS, Université de Strasbourg, 23 Rue du Lœss, BP 43, 67034 Strasbourg, France.*

[2] *Synchrotron SOLEIL, L'Orme des Merisiers, Saint-Aubin, BP 48, 91192 Gif-sur-Yvette, France*

[3] *Institut Jean Lamour UMR 7198 CNRS, Université de Lorraine, BP 70239, 54506 Vandœuvre les Nancy, France.*

[4] *IREPA LASER, Institut Carnot MICA, Parc d'innovation - Pole API, 67400 Illkirch, France*

[5] *Institut de Chimie, UMR 7177 CNRS, Université de Strasbourg, 4 Rue Blaise Pascal, CS 90032, 67081 Strasbourg, France.*

[6] *Instituto de Ciencia Molecular (ICMol), Universidad de Valencia, Catedrático Jose Beltrán 2, Paterna, 46980, Spain*

*: these authors contributed equally.
@: bowen@unistra.fr

**Abstract**

Recent theory[1–4] and experiments[3,5,6] have showcased how to harness quantum mechanics to assemble heat/information engines with efficiencies that surpass the classical Carnot limit. So far, this has required atomic engines that are driven by cumbersome external electromagnetic sources[3,5,6]. Here, using molecular spintronics[7], we propose an implementation that is both electronic and autonomous. Our spintronic quantum engine heuristically deploys several known quantum assets[1–6,8–11] by having a chain of spin qubits formed by the paramagnetic Co centers of phthalocyanine (Pc) molecules[12,13] electronically interact with electron-spin selecting Fe/$C_{60}$ interfaces[14]. Density functional calculations reveal that transport fluctuations across the interface can stabilize spin coherence on the Co paramagnetic centers, which host spin flip processes. Across vertical molecular nanodevices, we measure enduring dc current generation, output power above room temperature, two quantum thermodynamical signatures of the engine's processes, and a record 89% spin polarization of current across the Fe/$C_{60}$ interface. It is crucially this electron spin selection that forces, through demonic feedback and control, charge current to flow against the built-in potential barrier. Further research into spintronic quantum engines, insight into the quantum information processes within spintronic technologies, and retooling the spintronic-based information technology chain[7], could help accelerate the transition to clean energy.

**Main Text**

Classical engines convert heat into work by transferring heat from hot to cold thermal baths using a working substance (WS) that is sequentially put into contact with each bath. This upstream flow of heat thermodynamically increases the engine's entropy. During this process, nature limits the engine's maximum efficiency which cannot surpass an ideal value determined by the ratio of the temperatures of the two baths. This limit proven by Carnot in 1824 embodies the 2[nd] law of thermodynamics.

Quantum engines can surpass this limit by retooling its underlying concepts. Both theory[1–4] and experiments[3,5,6] suggest that additional work capacity, called 'ergotropy', can be harvested from quantum systems. In theory, the operation of these engine can be separated into *strokes* that emulate Nature's principle of least action[3]. A stroke's action is characterized by its duration and the rate at which it couples the WS to the baths. If the engine's baths and WS are endowed with quantum properties, they can constitute quantum assets (QAs)[1–4] that promote the storage and transfer of ergotropy.

So far, the only engines to implement a quantum advantage have coherently manipulated an atomic WS using 1-10MHz strokes [3,5,6] using microwave or laser external sources, i.e. the engine isn't autonomous. Conversely, despite much faster 2-10GHz electronic strokes, on-chip electronic engines[15,16] have so far exhibited sub-Carnot efficiencies due to mesoscopic WSs that are much more prone to quantum decoherence.

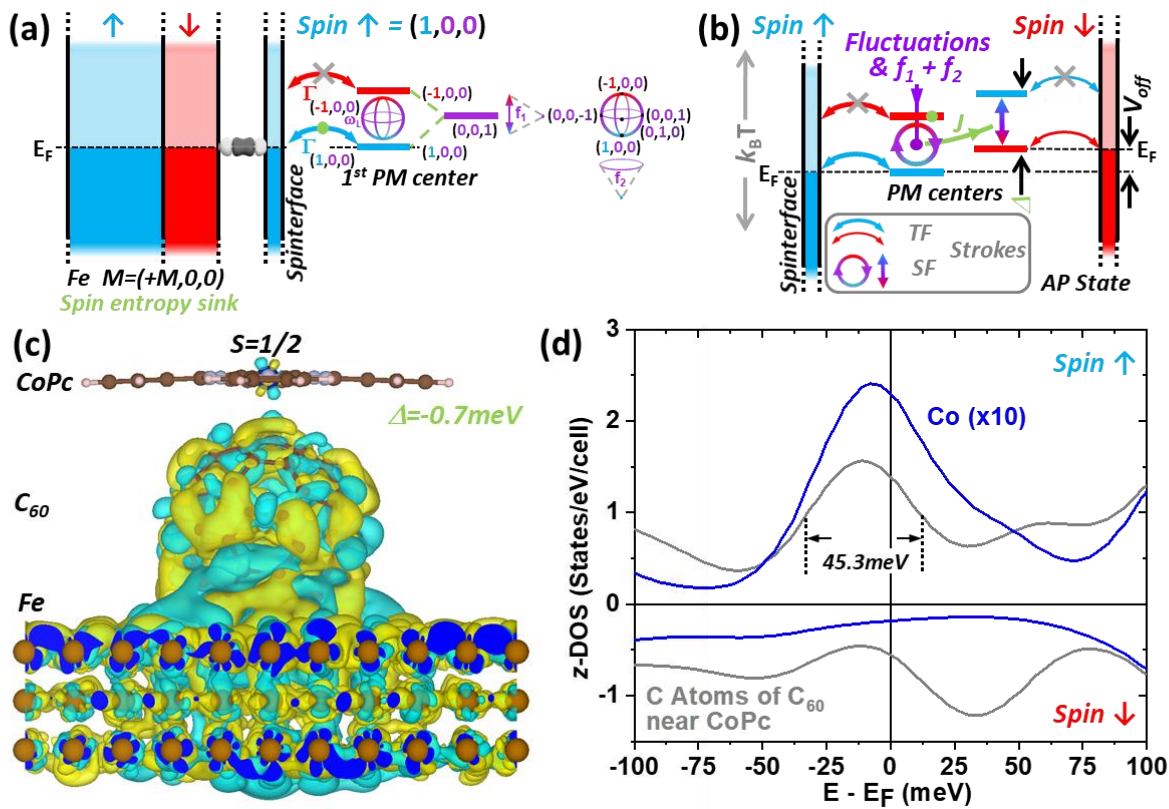

**Figure 1: a molecule-based spintronic quantum engine.** (a) Density of states schematic: spintronic implementation of the transport fluctuation (TF) stroke between the device electrode in its FM ground state and the spin states of the working substance (WS) 's nearest PM center, mediated by a spinterface with full transport spin polarization. Quantum coherence and decoherence processes on the WS are shown. This stroke appears in the (b) overall engine schematic that also shows the spin flip (SF) strokes on a PM center's spin states, and between the PM centers forming the WS, against thermal fluctuations for $k_BT > \Delta$. Quantum resources are color-coded in green in panels a-b. See main text for details. (c) Spatial charge transfer maps across Fe/$C_{60}$/CoPc reveal sizeable hybridization on $C_{60}$ (i.e. the spinterface) and electron tunneling between $C_{60}$ and CoPc across an antibonding state. Green/cyan isocontours depict charge gain/loss of 0.0007 e/Å$^3$. The AFM case is shown. (d) The antibonding state's DOS around $E_F$ reveals how the Co $d_z^2$ and C $p_z$ orbitals share a spectral feature that appears only in the spin ↑ band. This illustrates the high spin polarization and bandwidth of the TF stroke (see panel a).

We propose spintronics[7] as a potent scientific and industrial platform to achieve an electronic autonomous engine with a quantum advantage, and augment our prior spintronics-only description[17] with concepts drawn from quantum thermodynamics[1–6,8–11]. The system under study (see Methods) is a molecular device in which electronic interactions occur between the spin states of paramagnetic (PM) Co centers borne by phthalocyanine molecules (CoPc), which form the engine's WS, and those of ferromagnetic Fe electrodes (see Fig. 1a). Ultrathin $C_{60}$ intercalation layers ensure not only a spin selectivity[14] of these electronic interactions, but a partial magnetic decoupling[13] between Fe and Co spins, such that the Co spins may fluctuate in an effective magnetic field[17] that lifts the spin degeneracy by $\Delta$ (Fig. 1a-b). The resulting coherent superposition of spin states can constitute a QA (green dashed lines of Fig. 1a) if the spin population is inverted[3].

Our engine operates by harvesting the energy of these thermal fluctuations using electronic transport across the Fe/$C_{60}$ interface. Indeed, this so-called a *spinterface*[14] involves a low density of electron states with narrow energy width and high transport spin polarization. These properties enable the spinterface to electronically interact as a non-thermal bath[2,18,19] with the WS, i.e. constitute a QA (green dot in Fig. 1a). Furthermore, magnetic superexchange interactions[12] with energy J between the PM Co atoms of the WS promote a S = 1/2 molecular spin chain. This means that the spin coherence between the WS's spin states can be thermally driven through a magnetic phase transition, and dynamically modified by the electronic interaction with the non-thermal baths (Fig. 1c-d). This constitutes another QA[18]. In our design, the different thicknesses of the $C_{60}$ intercalation layers promote different transmissions of spin-conserved transport between the WS and each spinterface. This helps to break the detailed balance of charge transport fluctuations, leading to a preferred one-way direction of current. Overall, our engine's operation involving several QAs differs from the spin-based classical thermoelectric effects of spin caloritronics[35].

Thus, the quantum advantage we claim to observe heuristically originates from this tailored electronic interaction between the FM thin film and the discrete spin states of the PM centers. Our spintronic quantum engine transforms the thermal energy of spin fluctuations on the WS's PM centers into a directed dc current by rectifying the resulting transport fluctuations between the WS and the device electrodes. After presenting our experimental results, we will propose a more detailed description of the engine's operation.

**Transport Experiments**

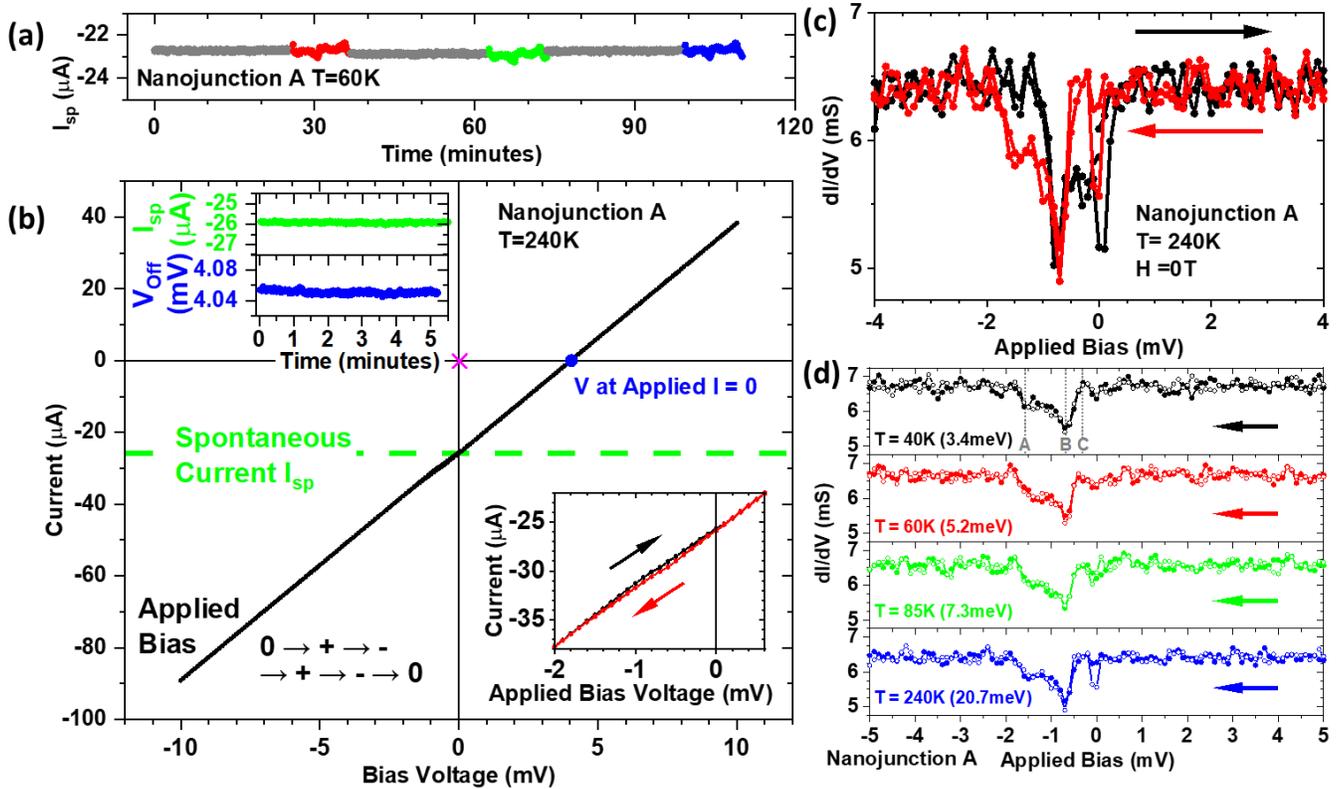

**Figure 2: Long-lived spontaneous electrical signals.** Data on metallic nanojunction A: (a) time dependence of $I_{sp}$ at $H = 0$ (grey) and upon applying a *H* field (red, green, blue) orthogonal to the electrode magnetizations. (b) *I(V)* data at 240K. The magenta crosspoint represents the experimental (*V,I*) error. The top insets show the time dependence of $I_{sp}$ and $V_{Off}$, while the zoom around *V* = 0 (lower inset) reveals an *I(V)* hysteresis that contains features with a sub-$k_BT$ spectral resolution in (c) the current derivative *dI/dV*. Forward (black) and return (red) traces are shown. (d) Return *dI/dV* traces for 40K, 60K, 85K and 240K reveal essentially identical features with a sub-$k_BT$ spectral resolution, i.e. the thermodynamical signature of a quantum asset.

Our nanoscale vertical junctions (see Methods) exhibit a large persistent non-zero spontaneous current $I_{Sp}$ ~-10 µA (see Fig. 2(a) for nanojunction A, and Suppl. Fig. S3 for other junctions) that is two orders of magnitude larger than the experimental offset (see Suppl. Note 2 on control experiments). Its amplitude is not strongly affected by intermittent sweeps of an external magnetic field up to 2T applied perpendicularly to the electrode magnetizations. This confirms that, in our implementation, the engine's primary energy source is not the external applied magnetic field[20]. We present in Fig. 2(b) repeated I(V) sweeps at 240 K, from which we infer a *slope resistance* $R_s$ = 157 Ω. From $I_{Sp}(t)$ (see top inset), we find at V = 0 that the *offset current* $I_{Off}$ = -26 µA = $I_{Sp}$. This suggests that, once a bias voltage *V* has been applied, and on the timescale of hours, applying *V* = 0 does not confer energy to the device. $I_{Off}$ and the *bias offset* $V_{Off}$ = 4.05 mV at $I$ = 0 (see top inset), are respectively 230x and 100x larger than the experimental offset errors observed for a 100 Ω calibrated resistance (magenta crosspoint in Fig. 2(b) and Suppl. Note 2).

The lower inset of Fig. 2(b) reveals a slight, hysteretic deviation from a linear response that depends on the sweep direction (red and blacks arrows). Within this 1.4 mV bias window, the numerical derivative (see Fig. 2(c)) reveals features with an energy width as low as 0.3 meV despite an expected thermal smearing of 2-3 $k_BT$ upon transport, with $k_BT$ = 20.7 meV here. This sub-$k_BT$ spectral resolution is mostly unchanged upon reducing thermal fluctuations by a factor of 6, as are the main spectral features (denoted A, B and C in Fig. 2(d)).

The persistence of these features with sub-$k_BT$ spectral width above the dissipation threshold induced by thermal fluctuations is most likely a result of feedback-induced noise reduction[21] and should therefore constitute the quantum thermodynamical signature of a non-thermal bath interaction within the engine operation. The 1.4 meV energy window in which the current-voltage characteristic deviates from a linear behavior pegs[17] a limit for the energy width of all the spin states along the WS's 3-spin long chain(s). As we will see, this is compatible with the spin splitting on each PM fluctuator $\Delta$ = 0.7 meV predicted by density functional theory (DFT) (see Fig. 1a-b-c).

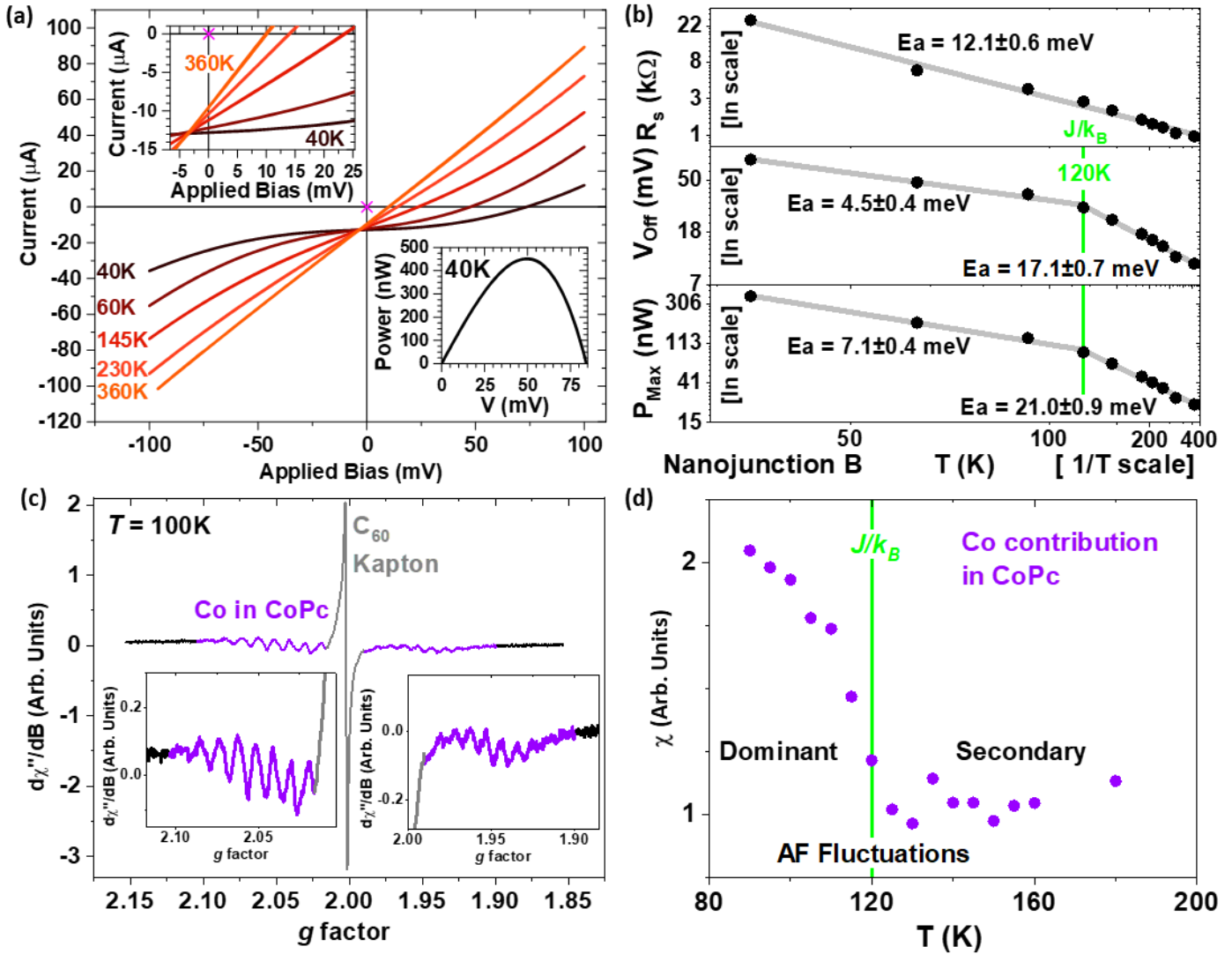

**Figure 3: Thermally activated electrical power across a magnetic phase transition.** (a) $I(V)$ data from nanojunction B at $H$ = 0T within 40 < $T$(K) < 360. Top inset: zoom at low bias. Lower inset: $P(V)$ data showing $P_{Max}$ = 450nW at 40K. The magenta crosspoint is the experimental error (see Methods). (b) ln vs 1/T plots of (top) $R_s$ around $V$ = 0, (middle) $V_{Off}$ and (bottom) $P_{Max}$. $P_{Max}$ decreases from 370nW at 40K to 24nW at 360K per two thermal activation regimes, with a 120K crossover temperature. The activation energy Ea is given for each regime. (c) EPR spectrum at T = 100K. The color-coded feature identification is described in Suppl. Note 3. Insets: zoom on the Co contribution from CoPc. (d) Temperature dependence of the Co contribution in CoPc to the EPR Intensity. A signal increase is observed for T < 120K.

Due to discrete states within the barrier[22], spintronic regimes involving multiple metallic and semiconducting transport nanochannels may coexist[23] in a device. The slight decrease in junction conductance with increasing temperature (see Fig. 2d) confirms the metallic nature of nanojunction A. This, and several other metallic nanojunctions, exhibit a *maximal output power* $P_{Max}$ such that $17 < P_{Max}$ (nW) $< 55$ for $150 < R_s$ (Ω) $< 800$ at 40 K. We now turn to semiconducting nanojunction B, for which we observe (see Fig. 3a) a mostly linear *I(V)* at 360 K ($R_s$ = 1.05 kΩ around V = 0) that becomes increasingly non-linear as *T* is lowered to 40 K ($R_s$ = 25.9 kΩ). The apparent common crosspoint of *I(V,T)* data (inset of Fig. 3a) is orders of magnitude beyond measurement artefacts (see magenta crosspoint and Suppl. Note 2) and might reflect a bias-induced symmetrization of the spin potential landscape against thermal broadening effects[17]. We observe $P_{Max}$ = 450 nW at 40 K (Fig. 3(a) inset). This represent a 450x increase over the previous record measured at 3 K and *H* = 1 T[20], and at 295 K a 270x improvement[17]. At 360 K, we still observe $P_{Max}$ = 24 nW, which is promising for applications. Our results thus not only dwarf those from possibly similar experiments[17,20,24], but also those from mesoscopic quantum heat engines[15].

As seen in the ln vs. 1/T plots in Fig. 3(b), the $R_s$ data follows $R_S = R_0 e^{-\frac{E_a}{k_B T}}$ with a single thermal *activation energy* $E_a$ over $40 < T$ (K) $< 360$. In contrast, $V_{Off}$ and $P_{Max}$ both exhibit two thermal activation regimes. We observe a ~3x increase in $E_a$ from the low T to the high T regime. This could reflect hopping transport[25] for T > 120K between spins onto and along the three member-long chain. As a corollary, coherent spin-polarized transport across the spin chain would occur for T < 120K. This suggests that the structural arrangement[12] of the CoPc spin chain in the junction's effective nanotransport path[26] yields a magnetic exchange energy J such that $k_B T_c$ = J with $T_c$ = 120K. Considering that $10 < T_c$ (K) $< 400$ theoretically[12], this is compatible with prior reports for Pb//CoPc ($T_c$ = *105K*) [27] and CoPc on Fe ($T_c \approx$ *72K*)[13] around the stable bulk α-CoPc phase ($T_c$ = *86K*)[12].

To independently determine $T_c$ for CoPc on $C_{60}$ in our nanojunction stack, we performed electron paramagmetic resonance (EPR) measurements on a $C_{60}$/CoPc multilayer stack (see Methods and Suppl. Note 3 for additional details). Referring to Fig. 3c, we observe a radical peak centered around the Landé g-factor *g* = 2. We also observe 8 oscillations on either side of the radical peak. Using reference samples and simulations (see Suppl. Note 3), we attribute the radical peak to Kapton and to $C_{60}$, and each set of 8 oscillatory features to the hyperfine structure of a paramagnetic center around a Co nucleus with nuclear spin 7/2. Upon subtracting the simulated Kapton and $C_{60}$ contributions from the main EPR dataset, we obtain the temperature dependence of the Co contribution in CoPc (see Fig. 3d). The Co ESR signal exhibits a clear increase for T < 120K, as expected when, below the magnetic phase transition at $T_c$, AF-correlated spin fluctuations along an spin chain of odd length dominate thermal disorder. As a quantum thermodynamical signature, we thus confirm that the magnetic phase transition of the WS plays a role in the spintronic engine's performance, as expected of this QA[18].

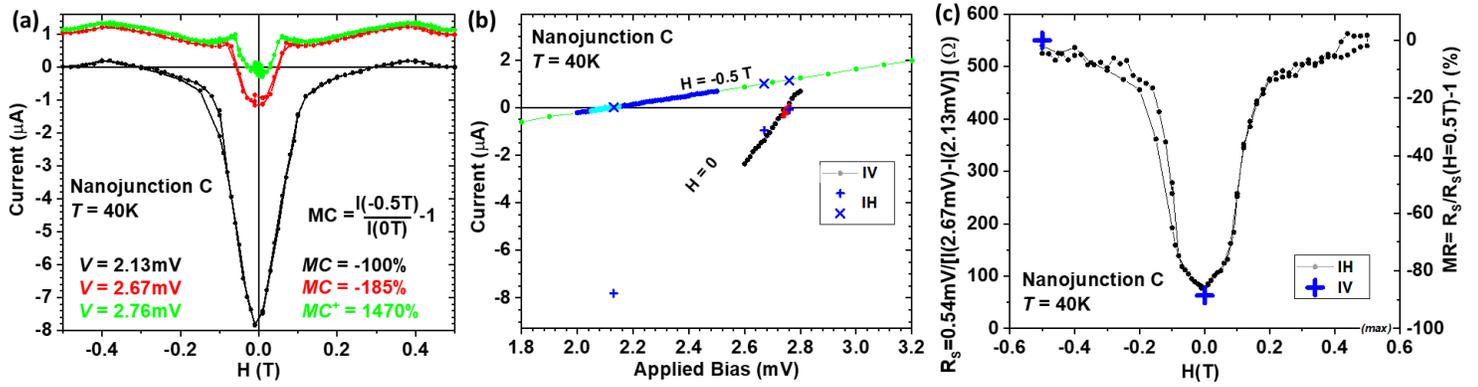

**Figure 4: Spintronic features of the thermal energy harvester.** (a) *I(H)* data acquired on nanojunction C at 40K. (b) I(V) data at *H*=0T and *H*=-0.5T, revealing a linear behavior with $R_S$=63Ω and $R_S$=550Ω around $V_{Off}$=2.13mV and $V_{Off}$ =2.76mV, respectively. The blue crosses reflect the *I(H)* data from panel a. (c) $R_S(H)$ calculated from two $I(V_0,H)$ datasets from panel (a). The blue crosses indicate $R_S$ inferred from panel b. The two spintronic $V_{Off}$ lead to extremal values *MC* (%)in *I(H)* data, as do the two spintronic $R_S$ regarding -100 < MR (%) < ∞.

To confirm these devices' spintronic underpinnings, the *I(H)* data at 40 K from nanojunction C in Fig. 4a reveal a strong in-plane *H* dependence of junction current, which saturates for |*H*| > 0.5 T (see Suppl. Note 4) upon achieving the parallel orientation of FM electrode magnetization. The current can be suppressed at *H* = 0 and *H* = -0.5 T for $V_{Off}$ = 2.76 mV and $V_{Off}$ = 2.13 mV, respectively. Thus, the junction's two magnetic states promote differing $V_{Off}$. They also drive a sign change in current at *V* = 2.67 mV, with *I* = ±1 µA. These *(V,I)* pairs lie well beyond possible experimental offsets (see Suppl. Note 2). The *I(V)* data of Fig. 4(b) confirms these (V,I) pairs. From $R_S$ (0T) = 63Ω and $R_S$ (-0.5T)=550Ω, we observe a magnetoresistance $MR = \frac{R_S(0T)}{R_S(-0.5T)} - 1$ = -89% (bounded by -100% for full spin polarization of the current). Using *I(H)* data from panel (a), we plot $R_S(H)$ in Fig. 4(c), and again obtain MR=-89%. Consistency between these three data panels is visualized by blue crosses in panels (b) and (c). The MR=-89% translates into an 'optimistic' magnetoresistance $MR' = \frac{R_S(-0.5T)}{R_S(0T)} - 1$ = 770%. This implies[7] an average transport spin polarization *P* = 89.1% of the two Fe/$C_{60}$ spintronic selectors[14]. The magnetocurrent $MC = \frac{I(-0.5T)}{I(0T)} - 1$ reaches -100% and 1470% at each spintronic $V_{Off}$. This showcases this device class as a spintronically controlled switch of current direction and flow.

**Model**

In the context of these experimental datasets, we now propose a basic description of how quantum assets heuristically drive the spintronic engine's operation, and utilize density functional theory to support our claim of spinterface-driven spin coherence on the WS's Co sites, in view of the long-lasting dc currents that were observed. Referring to Fig. 1(a), the (+M,0,0) magnetic orientation of the Fe layer sets a spin referential for other electronic interactions, starting with the spin-polarized charge transfer toward adjacent $C_{60}$ molecules. This generates the spinterface electronic state described previously, which can be seen as a magnetic quantum dot[17]. Prior to any electronic interaction, the Co on the phthalocyanine molecules bear energetically degenerate S=1/2 spins on the $d_{z^2}$ orbital ( (x,y,z)=(0,0,±1) on the Bloch sphere, see Fig. 1a).

The electronic interactions between the FM electrode and PM centers across the spinterface lift this degeneracy by Δ and describe the engine's spin-conserved transport fluctuation (TF) stroke. The forward TF stroke from the FM electrode to the nearest PM center on the chain is mediated by the spinterface's full spin polarization along the (1,0,0) direction.

This imposes the injection into the WS of coherent spins along (1,0,0) acting, against the WS's decoherence processes (in purple in Fig. 1a-b).

The reverse TF flow of current also requires that the spin on the WS be (1,0,0). This not only constrains the possible channels of spin decoherence for the WS's spins (Fig. 1a), but also describes an autonomous measurement of the WS state, i.e. a QA[4]. The spinterface acts as an autonomous Maxwell demon[28] that gains information on the WS and uses it as feedback to control the electron transfer rate across the barrier. Overall, these TFs generate resistive losses due to spin scattering between the spinterface and the FM film with very different spin polarizations. This rise in spin-based entropy of the FM essentially embodies the Landauer principle[29] : the quantum information gained by the measurement of the WS's spin state[4] is erased as the FM thin film scatters the excess spin. We believe that, importantly, the Fe FM thin film synergistically sinks this spin entropy by returning to the stable FM ground state with lower spin disorder[30].

To characterize our experiment's TF stroke, we computed the electronic interactions (see Methods) across the *bcc* Fe(110)/$C_{60}$(1ML)/CoPc(1ML) system (ML = monolayer). We observe a charge transfer of 1.231e$^-$ from Fe that is delocalized onto $C_{60}$ (Fig. 1c). A much weaker 0.03e$^-$ transfer from $C_{60}$ to CoPc occurs across an antibonding state. This tunneling-mediated electronic interaction results in spectral features across $E_F$ that are shared both by the Co $d_{z^2}$ orbitals of CoPc and the $p_z$ orbitals of the neighboring C sites of $C_{60}$ only for spin ↑ electrons (Fig. 1d)[a]. These interactions result in a lifting of the Co spin degeneracy by $\Delta$=0.7meV according to our DFT calculations (Figs. 1b-c). This effective magnetic field $H$ = 6T, generated by spintronic anisotropy[17], is much stronger than that for a noble metal spacer[31] because it originates from $C_{60}$-mediated Fe-Co antiferromagnetic superexchange[32].

The TF strokes quantum correlate the spinterfaces with the endmembers of the WS's spin qubits borne by 1-D molecular chains, thereby imposing boundary conditions on the WS. Each member of the spin chain is coupled to its two neighbors by magnetic exchange with characteristic energy $J$[12]. The WS has an antiferromagnetic ground state that is promoted by this coupling, and is reinforced when the two FM electrodes are antiparallel-oriented. Flipping one spin on the chain promotes an excited state that contains ergotropy. Since the WS is a quantum system with a reduced number of discrete energy levels, it can therefore exist in quantum coherent superposition of states[4,11], and its temperature is not well defined when isolated. The WS is coupled to the Fe electrode via the spinterface acting as a non-thermal bath with temperature $T$. The 45.3meV width of the electronic interaction (Fig. 1d) suggests that at least these endmembers remain quantum coherent with the spinterfaces up to $k_BT$ =45.3meV. This would explain the persistence of power output for $k_BT > J$, in line with the presence of quantum correlations above the critical point of a 2$^{nd}$ order phase transition. Indeed, the loss of correlations in the spin fluctuations along the molecular chain for $k_BT > J$ is valid for a macroscopic, statistical ensemble (e.g. in our EPR experiment). However, these correlations may persist within a single chain along the device's nanotransport path[13,26]. Furthermore, when $k_BT < J$ and T is further lowered, an additional source of coherence is provided by the increased predominance, against thermal energy, of the antiferromagnetic interaction between the WS's spin qubits.

During the engine's so-called *spin flip* (SF) stroke, the thermal fluctuations on and between the PM centers supply magnetic energy to the WS (i.e. charging), which will be extracted during the TF stroke *via* an ergotropic return from a non-passive (i.e. excited) state to the ground state[28]. Indeed, although stochastic in origin, these fluctuations can provide QAs by inducing coherence[1] and by energetically pumping the superposition of quantum states[3]. SF operations between qubits[4,5] are also responsible for flipping the spin on the chain endmember prior to the TF stroke (e.g. from (1,0,0) to (- 1,0,0) in Fig. 1a). The TF stroke thus drives electronic transport processes across the device, and thermalizes the WS with the spinterface non-thermal baths. This spintronic passivation effect enables the extraction of thermal energy and causes a directed current to flow. This asymmetric regime of operation is enabled by the structural/electronic asymmetries of our device, in this case thanks to different $C_{60}$ thicknesses (thick and thin transmission arrows in Fig. 1b). This generates different TF stroke frequencies and different spin-splitting values $\Delta$ on each PM center. This asymmetrizes the SF stroke frequency along the WS's qubits.

---

[a] The antiferromagnetic superexchange interaction between Fe and Co mediated by the $C_{60}$ molecule can be seen as two spinterfaces in series. This cancels the usual switch in sign of spin polarization across a spinterface[14].

To provide a basic justification for this engine's operation, we examine the frequency of the TF and SF strokes against the WS's spin-charge and spin-lattice decoherence processes[33], which occur along (0,±1,±1) (in purple in Fig. 1a-b). For CoPc, the respective frequencies for CoPc at 7 K are $f_1$ = 1 kHz and $f_2$ = 1 MHz [33], noting that $f_2$ can increase ~100x at 300 K[34]. For $k_BT \geq \Delta$, and given $\Delta$ = 0.7 meV, the SF stroke cutoff $f_{SF}$=169.5 GHz. Finally, according to the Heisenberg uncertainty principle, the 45.3meV minimum energy width of the TF stroke's spectral features (Fig. 1d) corresponds to an attempt frequency $f_{TF}$ = 1.38x10$^{14}$ Hz. This suggests[b] that $f_1$, $f_2$ << $f_{TF}$, $f_{SF}$ throughout our experiments at 40 < T (K) < 360, which would provide the necessary speed of the engine's cycle to overcome all decoherence processes, even at room temperature. As a QA[9], the TF stroke implements the quantum Zeno effect during the engine operation, i.e. it repeatedly sets and measures the WS's quantum thermodynamical state. This ultrafast dynamic stabilizes the spin chain against decoherence.

Our results introduce spintronics onto the quantum technologies roadmap as a very promising platform to implement autonomous quantum engines, and raise interesting questions for further research. A current can drive the electronic properties of paramagnetic centers: temperature, exchange coupling and even entropy production close to a phase transition without a temperature gradient [17,36,37]. Does this engine require electrical priming to operate, i.e. is it a quantum battery[6,9]? Can precise thermometry confirm the entropy sinking[30] by the FM state of the electrodes? Can certain aspects such as TF stroke asymmetry[10], exchange coupling[12] or number/parity of qubits be optimized, e.g. using molecular engineering? Does zero-point energy play a role? Answering these questions, for example using *in operando* electron spin resonance, would shed insight into the autonomous engine's multiple interlocking quantum resources, and help to determine its efficiency. Looking ahead, the most promising vector to industrialize this quantum technology could be MgO spintronics[17], which so far has mostly targeted information storage and processing needs[7]. To transform it into a dual-use information/energy technology workhorse will require mastering the insertion of paramagnetic centers[17], most likely through the control of oxygen vacancies[22]. Harvesting/storing this most basic form of energy --- ambient thermal energy, could help to alter our nomadic energy needs and accelerate the transition to clean energy.

**Methods**

Device preparation

Si/SiOx//Cr(5)/Fe(50)/$C_{60}$(*n* ML)/CoPc(3ML)/$C_{60}$(5ML)/Fe(10)/Cr(50) heterostructure stacks were grown in-situ and at room temperature in an ultra-high vacuum multichamber cluster by *dc* sputtering (metals) and thermal evaporation (CoPc). All numbers are in nm; 1ML $C_{60}$ = 0.9 nm. 1ML CoPc = 0.4 nm. The SiO$_x$ substrate was annealed at 110°C and allowed to cool down prior to deposition. Metals were sputtered in an Ar pressure $P$ = 1.5x10$^{-3}$ mbar (Cr) and P= 6x10$^{-4}$ mbar (Fe). Molecules were thermally evaporated under $P$ = 3x10$^{-9}$ mbar. The $C_{60}$ thickness *n* for nanojunctions A, B and C was 3 ML, 1 ML and 1 ML, respectively. The Fe layers are in-plane magnetized. Nanojunctions were crafted[13] using 300 nm-diameter SiO$_2$ nanobeads thanks to a recently developed resist- and solvent-free nanojunction process, and were wirebonded to a sample chip. The positive contact was connected to the junction's top electrode. All data were acquired with the sample at a constant, nominally uniform temperature $T$.

---

[b] We suppose here that the slower TF and SF strokes at the thicker $C_{60}$ layer also respect this inequality.

EPR Sample preparation

A Kapton(25μm)//C$_{60}$(6ML)/[ CoPc(3ML)/C$_{60}$(6ML)]$_{300}$/C$_{60}$(18ML)/MgO(30nm) sample was grown under UHV conditions. The Kapton substrate was annealed at 110°C and allowed to cool down prior to deposition. MgO was rf-sputtered from a MgO target at a Ar pressure of $1.5 \times 10^{-3}$ mbar. Molecules were thermally evaporated under $P = 3 \times 10^{-9}$ mbar. Under ex-situ conditions, the sample was cut into strips and inserted into a quartz tube. After flushing with He, the tube was sealed at a He pressure of 500mbar and inserted into the EPR cryostat. See Suppl. Note 3 for more experimental details.

Ab-initio calculations

The structural and magnetic properties of the *bcc* Fe(110)/C$_{60}$/CoPc system were calculated using density functional theory (DFT) by means of the *Vienna ab-initio simulation package* (VASP) and its built-in *projector augmented wave* (PAW) pseudopotentials. The exchange and correlation potentials are within the *generalized gradient approximation* (GGA) as parametrized by Perdew, Burke, and Ernzerhof. The van der Waals (VdW) weak interactions were computed within the GGA-D3 approach developed by Grimme and later implemented in the VASP package. A kinetic energy cutoff of 450 eV was used for the plane-wave basis set. To include the correlation effects of transition-metal 3$d$ electrons, we adopted the DFT-GGA+U method, using a Hubbard U-term of 5 eV and an exchange parameter of 1 eV. See references in Ref. [13] for further details. The supercell geometry comprises the following three Bravais vectors: (22.96 Å, 0, 0), (14.35 Å, 20.29 Å, 0), and (0, 0, 37 Å), using 80 iron atoms per layer in the *bcc* structure along (110). The vacuum region separating the periodic images is 25 Å.

Atomic positions were relaxed by annulling the force on the atoms to within $10^{-4}$ meV/ Å. First, using several structural models, we relaxed the subsystem composed of 3 layers of Fe(110) and C$_{60}$ to find the equilibrium distance. Then, using several structural models, we optimized the CoPc - C$_{60}$ distance to find the ground state. We then used these configurations to study the magnetic state of the entire Fe(110)/C60/CoPc system. To this end, we considered both ferromagnetic and antiferromagnetic couplings of the CoPc with Fe(110) substrate. After the atomic relaxations, we found that the antiferromagnetic configuration is more stable by about -0.7 meV. The C$_{60}$ - Fe(110) distance is about 2.26 Å, and that between C$_{60}$ and CoPc is 2.6 Å . In both magnetic states, the magnetic moment of the iron atoms below the C$_{60}$ molecules is 2.29 μB, and 2.72 μB far from C$_{60}$. The magnetic moment of Co in CoPc is +1.34 μB (FM coupling) and -1.34 μB (AFM coupling). The magnetic moment on the CoPc ligands is -0.082 μB (FM coupling) and +0.089 μB (AFM coupling).

**Acknowledgements**

We thank J. Arabski for invaluable support with sample growth, R. Whitney, Ph. Turek and Y. Henry for stimulating discussions, B. Doudin for collaborating on nanobead processing, the remaining members of the STNano technological platform staff for technical assistance with certain processing steps, and the IPCMS machine shop for support[38]. Work was performed using Synchrotron SOLEIL beamtime proposals 20170317 and 20180169. We acknowledge financial support from the Region Grand Est and Synchrotron SOLEIL, from CEFIPRA grant 5604-3, from the ANR (ANR-06-NANO-033-01, ANR-09-JCJC-0137, ANR-14-CE26-0009-01, ANR-21-CE50-0039), the Labex NIE "Symmix" (ANR-11-LABX-0058 NIE), the ITI QMat "SpinDrive", the EC Sixth Framework Program (NMP3-CT-2006-033370), the Contrat de Plan Etat-Region grants in 2006 and 2008, by « NanoTérahertz », a project co-funded by the ERDF 2014-2020 in Alsace (European Union fund) and by the Region Grand Est through its FRCR call, by the impact project LUE-N4S part of the French PIA project "Lorraine Université d'Excellence", reference ANR-15IDEX-04-LUE and by  the «  FEDER-FSE Lorraine et Massif Vosges 2014-2020 ", a European Union Program.

**Authors Contributions**

# Supplementary Information

**Quantum advantage in a molecular spintronic engine that harvests thermal fluctuation energy**

B. Chowrira[1,2*], L. Kandpal[1*], M. Lamblin[1], F. Ngassam[1], C.-A. Kouakou[1], T. Zafar[1], D. Mertz[1], B. Vileno[5], C. Kieber[1], G. Versini[1], B. Gobaut[1], L. Joly[1], T. Ferté[1], E. Monteblanco[3], A. Bahouka[4], R. Bernard[1], S. Mohapatra[1], H. Prima Garcia[6], S. Elidrissi[6], M. Gavara[6], E. Sternitzky[1], V. Da Costa[1], M. Hehn[3], F. Montaigne[3], F. Choueikani[2], P. Ohresser[2], D. Lacour[3], W. Weber[1], S. Boukari[1], M. Alouani[1], M. Bowen[1@]

## Note 1: Sample and cryostat details

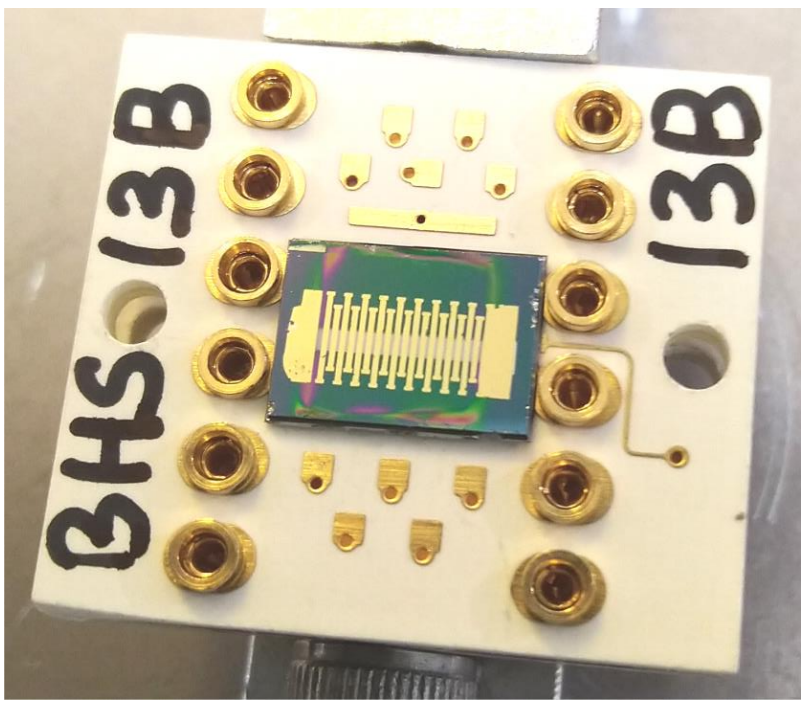

*Suppl. Fig. S1:* Photo of molecular junctions on a sample, mounted on a measurement chip.

We show in Suppl. Fig. S1 a typical photo of a sample prior to wirebonding. All junctions share a common horizontal electrode that is intersected by a vertical electrode for each junction. The nanopillars(s) within the areal overlap between top and bottom electrodes define the effective junction area, thanks to resist- and solvent-free nanobead processing[1]. We show in Suppl. Fig. S2(a) a corresponding photo after wirebonding and experiments. Panels (b-f) show additional information on the chemical composition of the sample.

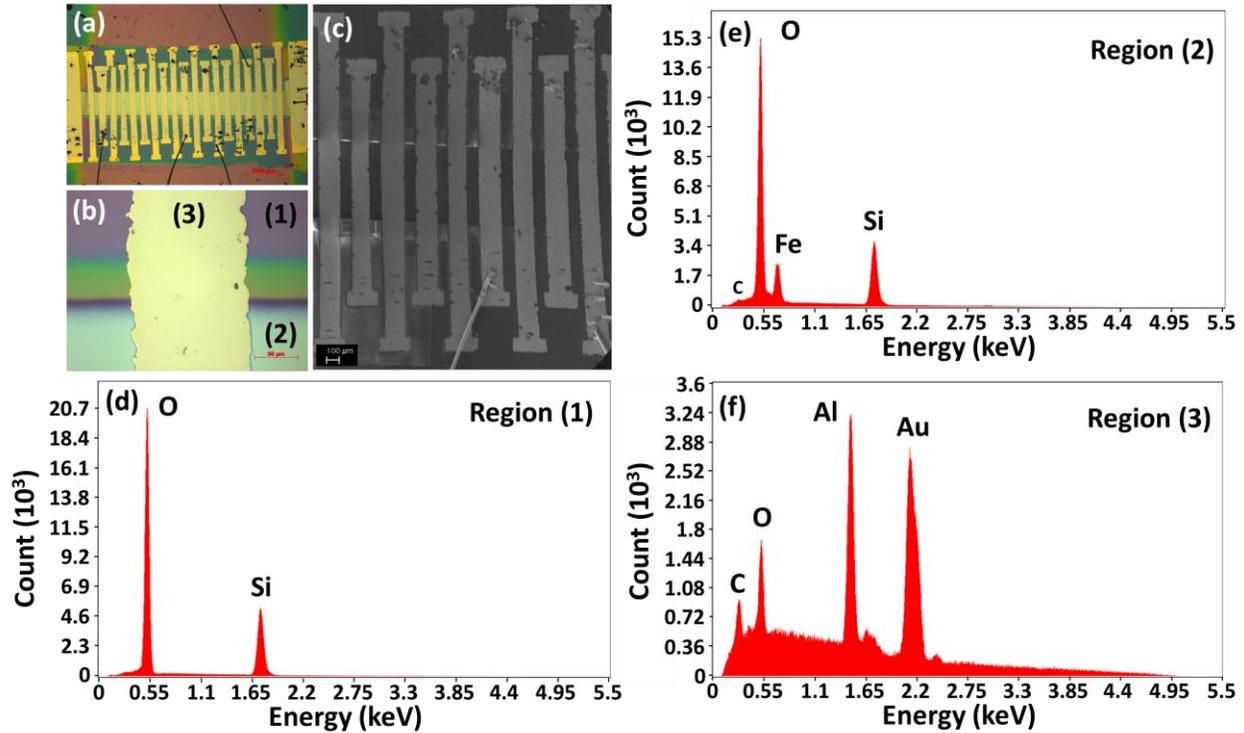

*Suppl. Fig. S2:* Topo-chemical analysis of a sample after wirebonding and measurement. (a) Optical image. (b) Zoom on the SiO$_2$ substrate with SiO$_x$ encapsulating layer (Region 1), on the SiO$_x$ covering the lower electrode (Region 2) and on the top technological contact (Region 3). (c) Scanning electron microscopy image. (e) EDX scan of Region 1 showing the presence of Si and O. (e) EDX scan of Region 2 showing the presence of Si, O and Fe from the lower electrode. (f) EDX scan of Region 3 showing the presence of Si and O, and of Al and Au from the top technological contact. All EDX scans were acquired at 5keV.

The static I-V, R-H or P-T device electrical measurements were performed using a high-end source-measurement unit (SMU) (Keithley 2636B) connected to a variable temperature insert using shielded cables. The electrical measurement wires inside the insert were thermally anchored to the insert to prevent thermal gradients on the sample. Unless otherwise stated, the SMU was placed in voltage source mode and a nominal *dc* voltage was applied. Out of 193 junctions processed, 10 were neither open-circuit or short-circuit, and 8 exhibited a sizeable combined current/voltage offset.

# Note 2: Discussion on possible artefacts

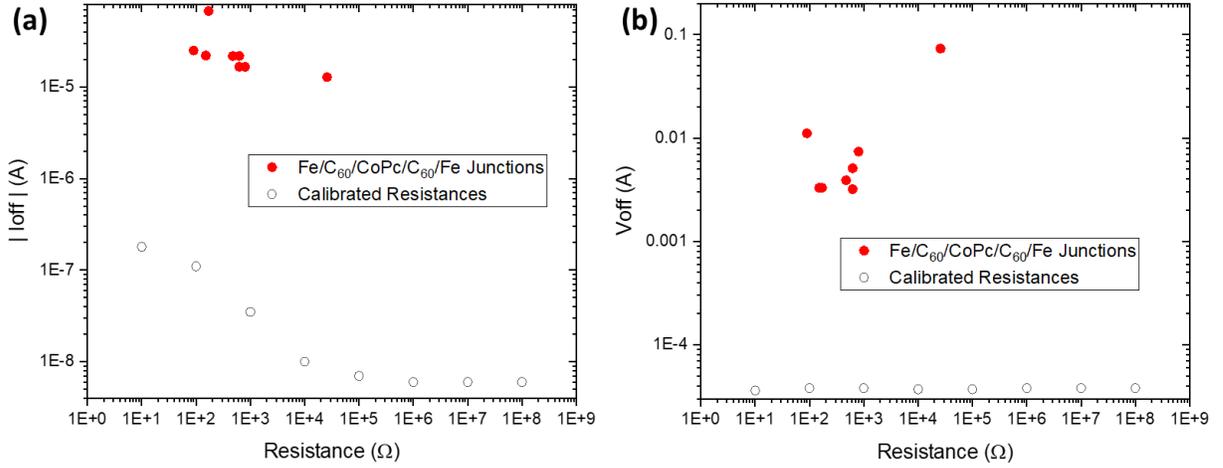

*Suppl. Fig. S3:* (a) Offset current and (b) offset voltage vs. resistance for Fe/$C_{60}$/CoPc/$C_{60}$/Fe nanojunctions (red circles) and calibrated resistances (open circles).

Using this cryostat, we did control measurements using calibrated resistances to determine current and voltage offsets. Suppl. Figure S3 shows these offsets measured on our working junctions (red circles) and on calibrated resistances (open circles) as a function of resistance. The junctions exhibit a combined current/voltage offset that is more than two orders of magnitudes larger than that found for calibrated resistances.

Measurements tool place the dark using an electrically grounded cryostat and long metallic wires[2]. An external microwave excitation is thus unrealistic.

The change in sign of the current upon switching the orientation of electrode magnetization ( using a low external field *H*, see Fig. 4(a)) and the voltage amplitude (up to 74 mV, see Fig. 3(b)) both point against a thermovoltage drop along the leads. The effect persists at *T* = 40 K when the sample heater is turned off, and decreases with increasing temperature. This casts aside a black-body radiation effect, as well as any stray temperature gradient between the cryostat's cold finger and the sample.

On nanojunction B, current and bias offsets that are orders of magnitude beyond experimental error for those $R_S$ are observed at all temperatures, and reach (84 mV,-13.8 µA) at 40 K (see lower inset of Fig. 3(a)). Indeed, using calibrated resistances to determine experimental bias and current offsets, we obtain (1 kΩ, 20 µV, 30 nA) and (10 kΩ, 30 µV, 0.8 nA). These experiment offsets are dwarfed by the device offsets (compare magenta crosspoint in Fig. 3a and top inset with data). Since the low bias slope resistance $R_s$ decreases with increasing *T*, an experiment current offset would increase with increasing *T* if this was an extrinsic effect (see Suppl. Fig. S3a), but $I_{Off}$ follows the opposite trend (see Fig. 3(a)). This is a strong indication that the effect isn't extrinsic. The crossing of *I(V,T)* data within (-5.3 < *V* (mV) < -2.9, -12.5 < *I* (µA) < -13.6) (see top inset of Fig. 3(a)), which also exceeds experimental error, suggests a bias-induced adjustment of the spin state potential landscape[3] so as to compensate thermal broadening effects.

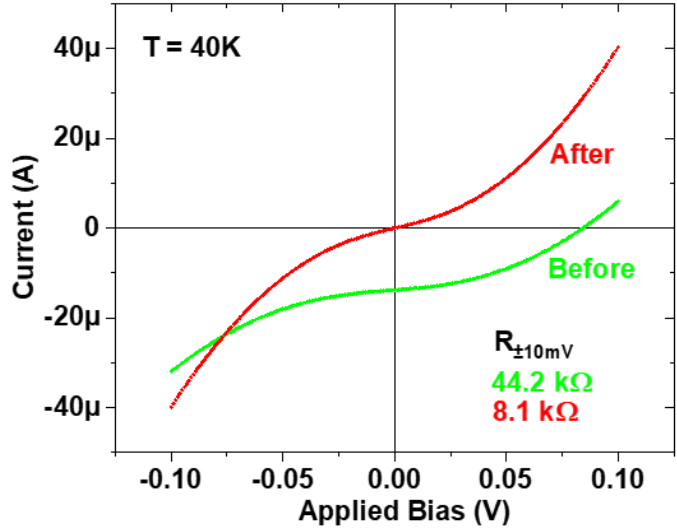

*Suppl. Fig. S4*: Junction degradation eliminates current offset. Measurements were performed on nanojunction B.

As electrostatic devices, junctions are subject to electrical degradation upon manipulation of experimental wiring. After one such manipulation on nanojunction B, we observe a reduction in device resistance, concurrently with a reduction in offset current to values within the systematic experimental error (see Fig. S4). This shows that our experimental setup isn't responsible for the current offset that is observed when the device isn't degraded.

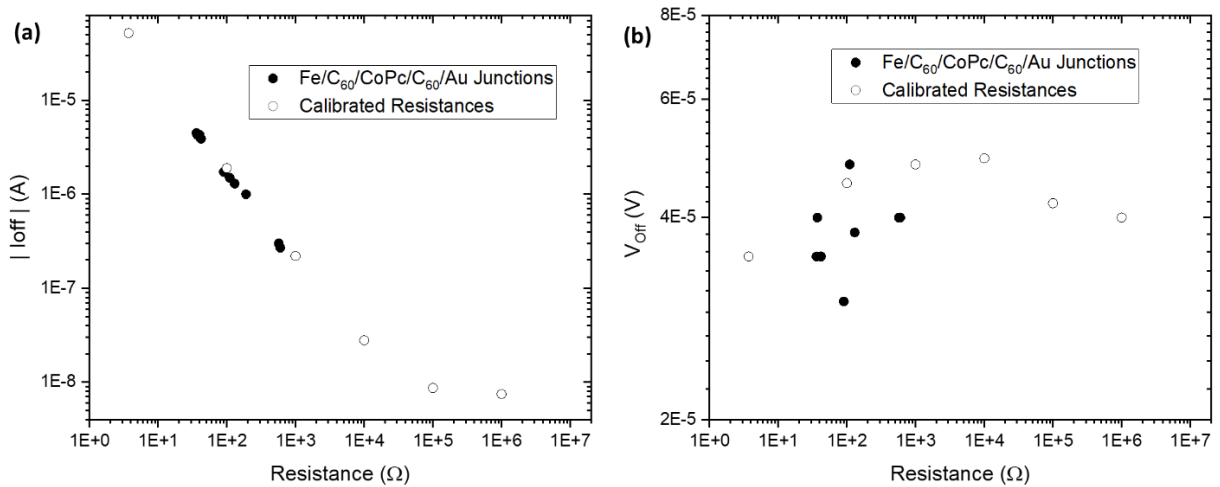

*Suppl. Fig. S5:* (a) Offset current and (b) offset voltage vs. resistance for Fe/$C_{60}$/CoPc/$C_{60}$/Au control nanojunctions (black circles) and calibrated resistances (open circles).

Control nanojunctions[1] containing only a CoPc spacer layer yielded current- and bias-offsets within the range found using calibrated resistances. We have also performed measurements on 9 control nanojunctions made from SiOx//Cr(5)/Fe(50)C$_{60}$(1ML)/CoPc(3ML)/C$_{60}$(5ML)/Au(10)/Cr(100) (numbers in nm) stacks, i.e. with only 1 ferromagnetic electrode. As with junctions for our main dataset, these control junctions also exhibited a temperature dependence of resistance that could be metallic or semiconducting. The latter behavior was observed on 4 junctions, in association with non-linear IVs. We plot in Suppl. Fig. S5 the offset currents (panel a) and offset voltages (panel b) for these control junctions, and compare them with those found using calibrated resistances. In contrast to our main dataset (see Suppl. Fig. S.3), we find essentially identical current and voltage offsets here.

## Note 3: Electron Paramagnetic Resonance Investigations

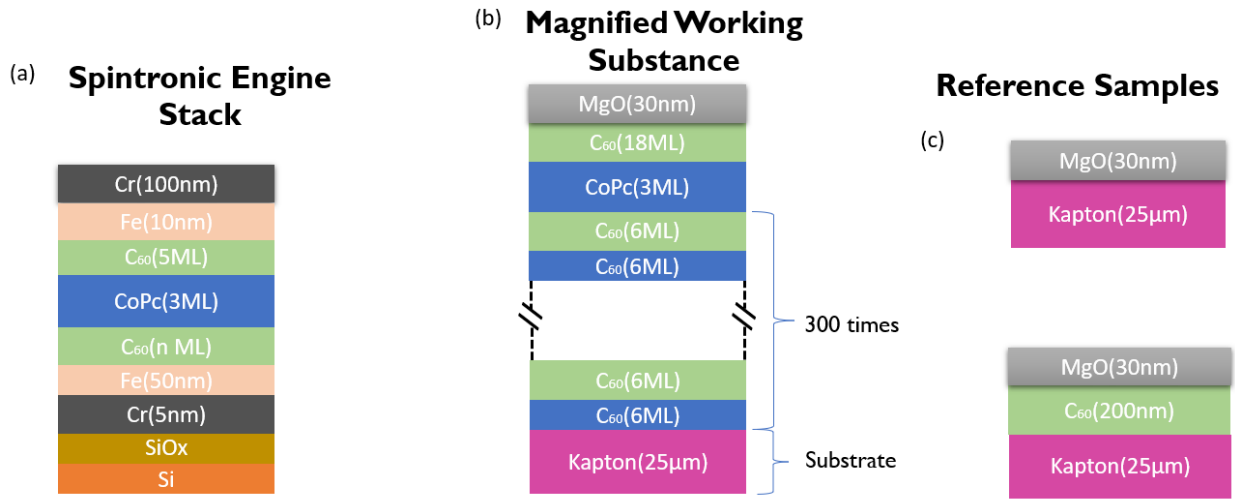

*Suppl. Fig. S6:* **Schematics of the sample stacks.** (a) Stack used to make nanojunction devices. (b) EPR Stack. (c) Reference stacks.

As previously mentioned, the spintronic engine is implemented experimentally using SiOx//Cr(5)/Fe(50)C$_{60}$(1-3ML)/CoPc(3ML)/C$_{60}$(5ML)/Au(10)/Cr(100) (numbers in nm) (Fig. S6a) stacks in which the working substance are the paramagnetic Co atoms of CoPc. As discussed by Serri et al[4], antiferromagnetic correlations in the paramagnetic fluctuations between Co spins along a structural CoPc molecular chain occur through 90° superexchange within and out of the molecular plane.

To measure the magnetic exchange energy between those paramagnetic centers, we used electron paramagnetic resonance (EPR) spectroscopy. EPR spectra were recorded with a continuous-wave X-band EMX-plus spectrometer (ca. 9.4 GHz, Bruker Biospin GmbH, Germany) equipped with a high-sensitivity resonator (4119HS-W1, Bruker Biospin GmbH). The g factor calibration was achieved in the experimental conditions by using a Bruker standard (strong pitch) with a known isotropic g factor of 2.0028. Samples were introduced into 4 mm outer diameter quartz tubes (Wilmad-Labglass) and sealed under 500mbar of He. A variable temperature was achieved by using a continuous-flow liquid-nitrogen cryostat. The main experimental parameter for the EPR stack were: microwave power 2.549 mW; modulation amplitude 2G; conversion time and time constant were setup at 48 and 20.48 ms, respectively; 500 G were swept in 5 min, and several spectra were accumulated to ensure a good signal-to-noise ratio (S/N). All experimental EPR spectra were analyzed through computer simulation using lab-made scripts based on the Easyspin toolbox[5] under Matlab (Mathworks) environment. Strain parameters on g, field (H) and hyperfine coupling (A) were used to account for the experimental line-broadening.

To measure the magnetic exchange energy between those paramagnetic centers, we used Electron Paramagnetic Resonance (EPR) spectroscopy. Since we seek to observe the signal of the paramagnetic center of the 3 monolayers of CoPc, our EPR sample cannot contain other strongly magnetic layers (Cr, Fe). To minimize the signal arising from the substrate, and to increase the Signal to Noise ratio (S/N), we used flexible 25 µm thick Kapton substrates[6]. To enhance the magnetic response of the 3 monolayers of CoPc encapsulated between $C_{60}$ monolayers and achieve sufficient EPR signal and S/N, we repeated this CoPc stacking on the sample. We thus made the "Magnified working substance" sample (Fig. S6b): Kapton(25µm)//$C_{60}$(6ML)/ [$C_{60}$ (6ML)/CoPc(3ML)]$_{300}$/$C_{60}$ (18ML)/MgO (30nm). This sample is denoted 'EPR stack' hereafter. Its thickness was confirmed using profilometry.

Attempting to compare magnetotransport and EPR experimental results involves the following assumptions:

1) since $C_{60}$ is amorphous, each 6ML of $C_{60}$ will provide the same surface for CoPc growth, such that the same CoPc structure, and thus CoPc spin chain properties, will be repeated.
2) these properties are those of the CoPc(3ML) ultrathin film grown atop $C_{60}$ in the transport heterostructure.
3) the CoPc properties are homonegeneous, so that a statistical EPR study will be representative of the properties in the junction's effective nanotransport path[7]
4) 6ML of $C_{60}$ are sufficient to magnetically decouple the CoPc layers.
5) in the spintronic engine stack, Fe does not alter the magnetic exchange energy along the spin chain.

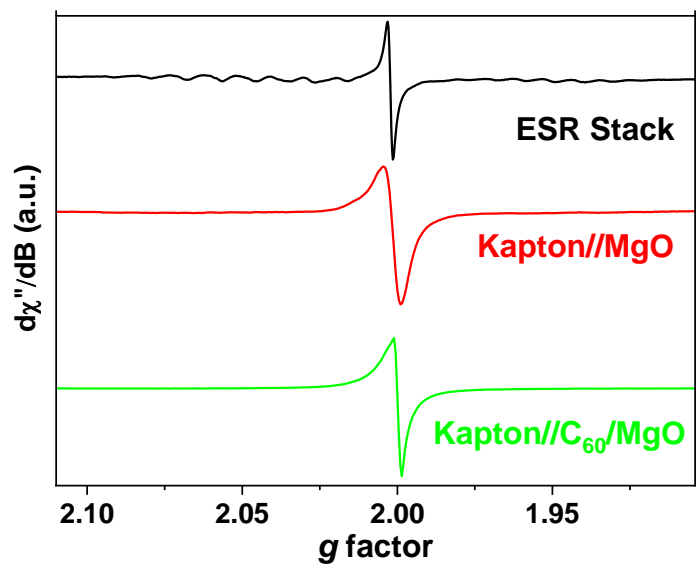

*Suppl. Fig. S7:* **Experimental EPR spectra** at T = 100K of (black) the 'EPR Stack', and of the (red) Kapton//MgO and (green) Kapton//C60/MgO references. The data are offset for clarity.

We presented in Fig. 3c the EPR signal for the EPR stack at T = 100K. To justify the attribution of spectral features therein, we first present in Suppl. Fig. S7 experimental spectra at T = 100K for our reference stacks. The Kapton // MgO and Kapton // $C_{60}$ / MgO stacks both provide a radical contribution at $g \sim 2$.

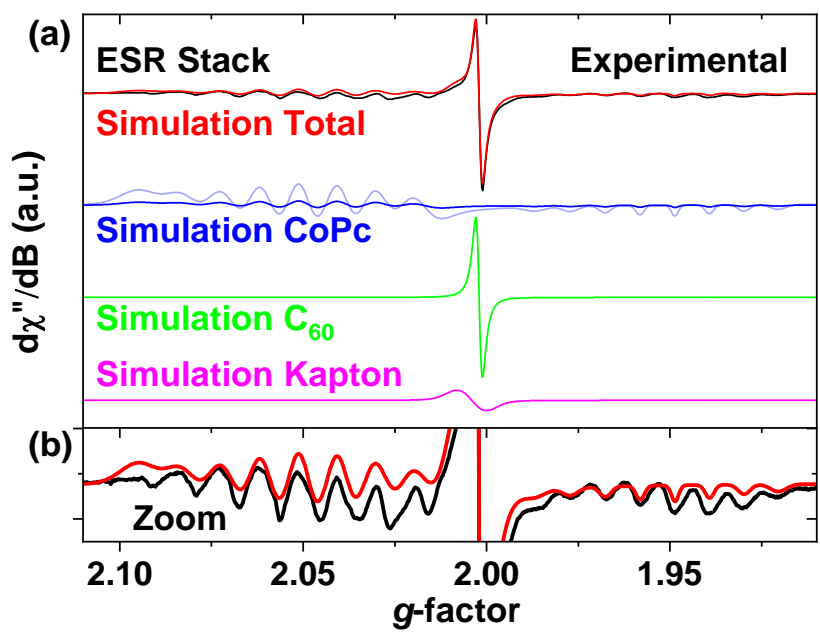

*Suppl. Fig. S8:* **Deconvolution of the EPR Stack spectrum.** (a) (Black line) Experimental data for the EPR Stack at T = 100 K. The fit (red line) was obtained using the simulated contributions of (blue line; transparent blue: 5 x zoom) Co in CoPc with > 80 % of the total spins. (Green line) $C_{60}$ and (purple line) Kapton together contribute <15-20% of the total amount of the probed spins. The data are offset for clarity. (b) Zoom on the data of panel a.

To confirm these feature identifications, and with background literature[8], we simulated the EPR stack experimental spectrum. An example at $T$ = 100K is shown in Suppl. Fig. S8. Reference spectra allow us to distinguish and deconvolute the radical contributions from both the Kapton (pink, g ~ 2.004) and $C_{60}$ (green, g ~ 2.002) radicals and the CoPc hyperfine fingerprint (blue). The Co nuclear spin I = 7/2 leading to 2I+1 = 8 hyperfine lines and the axial symmetry of the CoPc molecule point to distinct $g_\perp$ (at higher $g$) and $g_{//}$ (at lower $g$) features. Simulations led to the following numerical values: $g_\perp$ ~ 2.0055 and $g_{//}$ ~ 1.953 together with the corresponding hyperfine coupling constant: $A_\perp$ ~ 47.5 MHz, $A_{//}$ ~ 45 MHz.

The EPR intensity ($\chi$), proportional to the concentration of probed paramagnetic centers, is obtained through the double integration of the EPR signal (d$\chi$"/dB). This, together with qualitative simulations of the EPR spectrum (Suppl. Fig. S8), enables the deconvolution of the CoPc EPR Intensity from $C_{60}$ and Kapton radical contributions (Suppl. Fig. S9).

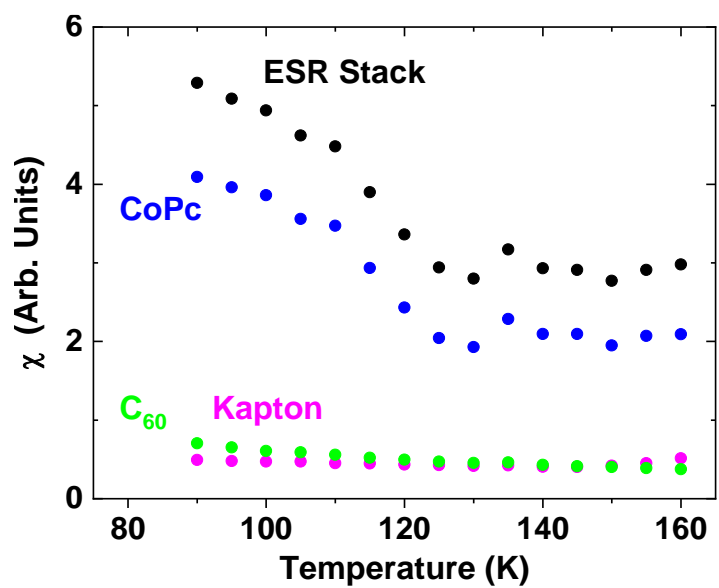

*Suppl. Fig. S9:* **Temperature dependence of the EPR Stack and its contributions.** Double integration of the (black) EPR Stack experimental spectra and of the (green) $C_{60}$ and (purple) Kapton simulated spectra. The blue data, attributed to CoPc, is obtained by subtracting the $C_{60}$ and Kapton contributions from the EPR Stack data.

We present in Suppl. Fig. S9 the temperature dependence of EPR intensity for our EPR stack. We observe that the EPR intensity exhibits a clear intensity increase below 120K. As expected, the $C_{60}$ and Kapton contributions represent an order-of-magnitude smaller integrated signal and do not exhibit a change around 120K. Subtracting these contributions from the experimental data yields the blue dataset, also shown in Fig. 3d of the main text. We note that the intensity increase below 120K arises in great part from a change in the background within the EPR spectrum range that contains the CoPc hyperfine structure. Furthermore, given our calibration, we estimate that the EPR intensity amounts to ca. a few % of the expected intensity. This could reflect inhomogeneities in our sample (cf. assumption 3 above) and/or dipolar effects. To conclude, our EPR experiments evidence the presence of a magnetic phase transition at T=120K within 3ML-thick layers of CoPc surrounded by $C_{60}$, *i.e.* that the spin chain's magnetic exchange energy J = $K_BT_C$ with $T_C$ being of *ca.* 120K.

**Note 4: Additional Data**

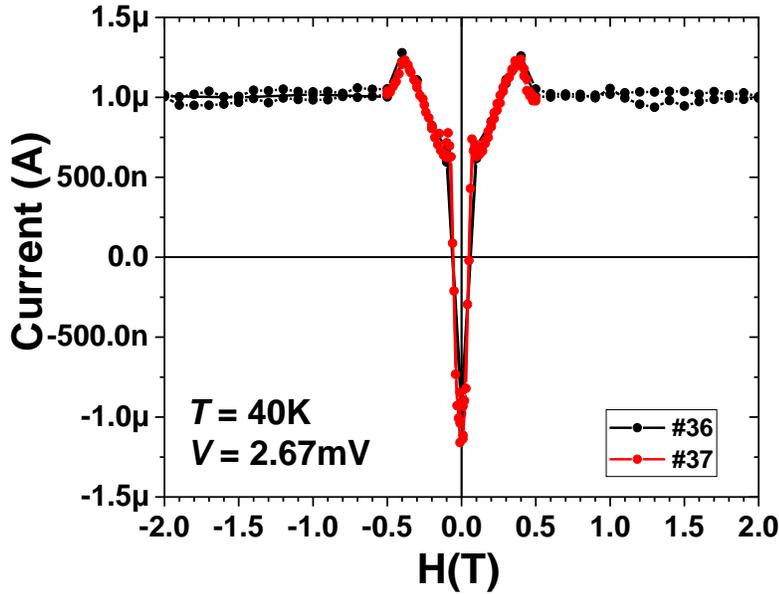

*Suppl. Fig. S10:* **Current-Magnetic field dependence up to ±2T.** Consecutive scans are shown.

We present in Suppl. Fig. S10 the field dependence of junction current on nanojunction C. We observe that the current is constant for 0.4 < |H(T)| < 2. We deduce that the parallel orientation of junction electrodes is achieved for |H| > 0.5T.

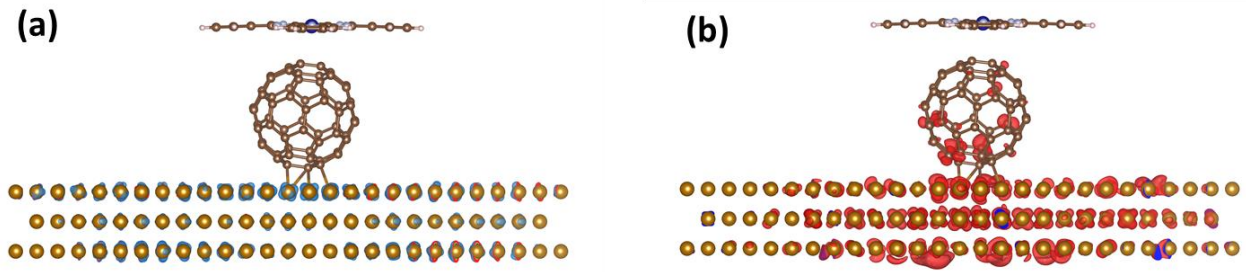

*Suppl. Fig. S11:* **Relaxed structural model of the *bcc* Fe(110)/C60/CoPc system.** The electron spin (a) down and (b) up density for in the energy range $E_F \pm 25$meV is shown.

We present in Suppl. Fig. S11 a schematic of the structural geometry of the bcc Fe(110)/C60/CoPc system that was used in the calculation of Fig. 1c. The flat asorption of CoPc on top of $C_{60}$ is similar to that found in the literature[9].